\documentclass[aps,pra,twocolumn,groupedaddress,showpacs,nofootinbib,floatfix]{revtex4-1}
\usepackage{graphicx}
\usepackage{amsmath}
\usepackage{dcolumn}
\usepackage{bm}

\begin{document}


\title{Diamagnetic correction to the $\bm{^9}$Be$\bm{^+}$ ground-state hyperfine
constant}
\thanks{Work of the U. S. government. Not subject to U. S.
copyright.}

\author{N. Shiga}
\email{shiga@nict.go.jp}
\altaffiliation{Present address: National Institute of Information and Communications Technology, Tokyo 184-8795, Japan}
\author{W. M. Itano}
\email{itano@boulder.nist.gov}

\author{J. J. Bollinger}

\affiliation{National Institute of Standards and Technology,
Boulder, CO 80305, USA}

\date{June 22, 2011}

\begin{abstract}
We report an experimental determination of the diamagnetic correction to the
$^9$Be$^+$ ground state hyperfine constant $A$. We measured $A$ =
$-625\,008\,837.371(11)$~Hz at a magnetic field $B$ of 4.4609~T. Comparison with
previous results, obtained at lower values of $B$ (0.68~T and 0.82~T), yields the
diamagnetic shift coefficient $k$ = $2.63(18) \times 10^{-11}$ T$^{-2}$, where
$A(B)=A_0\times  (1+k B^2)$. The zero-field hyperfine constant $A_0$ is
determined to be $-625\,008\,837.044(12)$ Hz. 
The $g$-factor ratio ${g_I}^\prime/g_J$ is determined to be
 $2.134\,779\,852\,7(10) \times 10^{-4}$,
which is equal to the value measured at lower $B$ to within experimental error. 
Upper limits are placed on some other corrections to the Breit-Rabi formula.
The measured value of $k$ agrees with theoretical estimates.
\end{abstract}


\maketitle

\section{Introduction\label{sec:intro}}

Transition frequencies between hyperfine-Zeeman sublevels of ground or
metastable electronic states of atoms can in some cases be measured extremely
accurately. One example is the ground-state hyperfine transition of
$^{133}$Cs, which is currently the basis of the SI second \cite{terrien68}.
Its frequency can be measured with a relative accuracy of better than $5 \times
10^{-16}$ \cite{parker10}. Because of the accuracy with which the energy
separations can be made in the ground states of Cs and other atoms, various
small contributions to the energies can be observed and compared with
calculations.

The relative energies of the sublevels of an atom with electronic angular
momentum $\textbf{J}$ and nuclear spin $\textbf{I}$, in a fixed magnetic field
$\textbf{B}$, are, to a good approximation, determined by the effective
Hamiltonian
\begin{eqnarray}
H &=& hA\textbf{I}\cdot\textbf{J} - \bm{\mu}_I\cdot \textbf{B} - \bm{\mu}_J\cdot
\textbf{B}\nonumber \\
 & = & hA\textbf{I}\cdot\textbf{J} +  {g_I}^\prime\mu_B\textbf{I}\cdot \textbf{B} +
 g_J\mu_B\textbf{J}\cdot \textbf{B}.\label{eq:BRHamiltonian}
\end{eqnarray}
On the right-hand side of Eq.~(\ref{eq:BRHamiltonian}), the first term is the magnetic
dipole hyperfine interaction, the second is the nuclear Zeeman interaction,
and the third is the electronic Zeeman interaction. Here, $h$ is the Planck constant, $A$ is the magnetic
dipole hyperfine constant, and $\bm{\mu}_I$ and $\bm{\mu}_J$ are the nuclear
and electronic magnetic moment operators, respectively. The $g$-factors are defined by
$g_J=-\mu_J/(J\mu_B)$ and ${g_I}^\prime = -\mu_I/(I\mu_B)$, where $\mu_B$ is
the Bohr magneton. The prime in ${g_I}^\prime$ is to distinguish it from the
alternative definition $g_I = \mu_I/(I\mu_N)$, where $\mu_N$ is the nuclear
magneton. For $I \ge 1$ and $J \ge 1$, other terms, such as the electric
quadrupole hyperfine interaction, should also be included on the right-hand side of
Eq.~(\ref{eq:BRHamiltonian}).

For $J=\frac{1}{2}$, the eigenvalues of $H$ are given analytically by the
solutions of quadratic equations. The expression for the energy eigenvalues is
known as the Breit-Rabi formula \cite{breit31,ramn56}. (In its original form,
the Breit-Rabi formula did not include the nuclear Zeeman interaction, since
it was considered to be negligible \cite{breit31}.) The Breit-Rabi formula for
the energies of the ($F$, $m_F$) sublevels in a state with $J=\frac{1}{2}$ and $I \ge \frac{1}{2}$ is
\begin{widetext}
\begin{eqnarray}
E(F,m_F)& = &h A
\left(-\frac{1}{4}+ \frac{{g_I}^\prime m_F \mu_B B}{hA} \pm \frac{2I+1}{4}\sqrt{1 + \frac{4 m_F}{2I+1}X + X^2}
\right)\nonumber\\
& = &h A
\left(-\frac{1}{4}+ \frac{2\gamma}{1-\gamma} m_F X \pm\frac{2I+1}{4}\sqrt{1 + \frac{4 m_F}{2I+1}X + X^2}
\right). \label{eq:BreitRabi}
\end{eqnarray}
For the special case $F=I+\frac{1}{2}$, $m_F=\pm (I+\frac{1}{2})$,
\begin{equation}
E(F,m_F) = hA\left[ \frac{I}{2} \pm \left(\frac{g_J}{2} + I{g_I}^\prime\right)\frac{\mu_B B}{hA}\right]
 = hA\left[ \frac{I}{2} \pm \frac{(1 + 2I\gamma)}{(1-\gamma)}X\right].
\label{eq:BreitRabiStretch}
\end{equation}
\end{widetext}
Here, $\textbf{F}=\textbf{J}+\textbf{I}$, and $m_F$ is the eigenvalue of $F_z$.
At finite $B$, the energy eigenstates are not eigenstates of $\textbf{F}^2$, except
for the $F=I+\frac{1}{2}$, $m_F=\pm (I+\frac{1}{2})$ states.  Nonetheless, we label them
by the value of $F$ that is valid at $B = 0$. Here  $X \equiv \mu_B B(g_J-{g_I}^\prime)/[(I+1/2)hA]$ is a
dimensionless quantity proportional to $B$, and $\gamma$ $\equiv$ ${g_I}^\prime/g_J$ is the $g$-factor ratio.  The $\pm$ sign in Eq.~(\ref{eq:BreitRabi}) corresponds to the states labeled by $F=I\pm \frac{1}{2}$. For the $1s^2 2s\, ^2S_{1/2}$ ground electronic state of $^9$Be$^+$, which was the subject of this study, $I=\frac{3}{2}$.
The value of $g_J$ for the ground electronic state of $^9$Be$^+$ has been determined by measuring the $^9$Be$^+$ cyclotron frequency and a hyperfine-Zeeman transition frequency at the same magnetic field \cite{wineland83}.  The value is $g_J = 2.002\,262\,39(31)$, calculated with the use of the best current value of the proton-electron mass ratio \cite{mohr08}. 
For $^9$Be$^+$, $X \approx -22.414 B(\text{T})$.

There are several ways in which the experimental energy separations
can deviate from those predicted by the Breit-Rabi formula. For $I>\frac{1}{2}$, it is
possible that, at a fixed value of $B$,  no values of the parameters {$A$, ${g_I}^\prime/g_J$, $X$}
can be found that are consistent with all of the measured
energy separations. It is also possible that the values of the parameters
determined at one value of $B$ are not consistent with those determined at
another value.

There are several possible sources of deviations from the Breit-Rabi formula.
If there is another electronic energy level that is close in energy,
hyperfine or Zeeman interactions can mix the electronic states.  For the
ground state of an alkali atom or an alkali-like ion, there are no nearby
electronic states, so such effects are small.  More important are diamagnetic
contributions to the interaction between the atom and the magnetic field that
are neglected in the effective Hamiltonian given by
Eq.~(\ref{eq:BRHamiltonian}). Diamagnetic corrections to the Breit-Rabi
formula were first considered by Bender \cite{bender64}. He calculated the
size of the deviation in the ground state of $^{133}$Cs to be equivalent to a
fractional shift in $A$ of $\delta A/A$ = $3.9 \times 10^{-10}$ $B^2$, where
$B$ is expressed in teslas. This effect, called the dipole diamagnetic shift
in atomic hyperfine structure, is due mainly to a magnetic-field-induced change in the
electronic spin density at the nucleus.

Measurements of magnetic-field-dependent deviations from the Breit-Rabi
formula in the ground state of Rb were made by Larson and coworkers
\cite{economou77,lipson86,fletcher87}. The dipole diamagnetic shift was
observed experimentally  in the hyperfine structure of $^{85}$Rb
\cite{economou77} and later in $^{87}$Rb \cite{lipson86}. A quadrupole
diamagnetic shift was observed in $^{85}$Rb and $^{87}$Rb  \cite{lipson86}. In
contrast to the dipole shift, the quadrupole shift can be thought of as a
magnetically induced electric quadrupole hyperfine interaction, which would be
absent in a pure $J=\frac{1}{2}$ state. The diamagnetic potential, which contains a rank-2 spherical tensor part, breaks the spherical symmetry, so that the electronic state is no longer an exact eigenvalue of $\textbf{J}^2$. The signature of the quadrupole diamagnetic term is an energy shift
proportional to  $[I(I+1)-3m_I^2]QB^2/[I(2I-1)]$, where $Q$ is the nuclear quadrupole moment.  In Rb, the
quadrupole shift is smaller than the dipole shift by about three orders of
magnitude. Another magnetic-field-dependent energy
term was observed in $^{85}$Rb and $^{87}$Rb \cite{fletcher87}. The term was explained by Fortson
\cite{fortson87} and is called the hyperfine-assisted Zeeman shift
\cite{harris88}. The shift of a level is proportional to $[{m_I}^2 m_J-I(I+1)m_J+m_I/2 ]{({g_I}^\prime})^2B$ and is due to mixing of
higher electronic states with reversed electronic spin into the ground electronic state
by the magnetic dipole hyperfine interaction.

The ground-state hyperfine constant of $^9$Be$^+$, $A$, was measured with a
fractional uncertainty of $2.4 \times 10^{-6}$ by Vetter {\em et al.} by
rf-optical double resonance \cite{vetter76}. The fractional uncertainty of $A$
was decreased to $1.6 \times 10^{-11}$ by Wineland {\em et al.}, in
measurements made with laser-cooled ions in a Penning trap \cite{wineland83}.
The low uncertainty was due mainly to the use of transitions for which the
first derivative of the frequency with respect to $B$ is zero. Nakamura {\em
et al.} measured $A$ with a fractional uncertainty of $1.2 \times 10^{-9}$ in
laser-cooled ions in a linear rf trap, at $B$ = 0.47~T
\cite{nakamura02}. Their value of $A$ differed from that of
Ref.~\cite{wineland83} by about two standard deviations. Okada et al. \cite{okada08} have measured $A$ for $^7$Be$^+$ in a linear rf trap.

Based on theoretical considerations and the experimental results for Rb, the ground-state hyperfine constant of $^9$Be$^+$, $A$, is assumed to have a weak quadratic dependence on $B$ such that $A(B) = A_0\times  (1 +kB^2)$. Transition frequencies measured at different values of $B$ are used to determine the diamagnetic shift coefficient $k$. 

The present experiment, on the measurement of the hyperfine-Zeeman transition
frequencies in the ground electronic state of $^9$Be$^+$ in a high magnetic field ($B$ = 4.4609~T), is described in
Sec.~\ref{sec:highfield}. In Sec.~\ref{sec:diamagnetic}, the high-field
results are combined with the previous, lower-magnetic-field measurements to
obtain a value for $k$. A theoretical estimate of $k$ is given in Sec.~\ref{sec:theory}.

\section{\label{sec:highfield}High-Field Experiment}

\subsection{\label{subsec:energylevels}Atomic energy levels and transitions}

\begin{figure}[tbhp]
\begin{center}
\includegraphics[width=8.2cm]{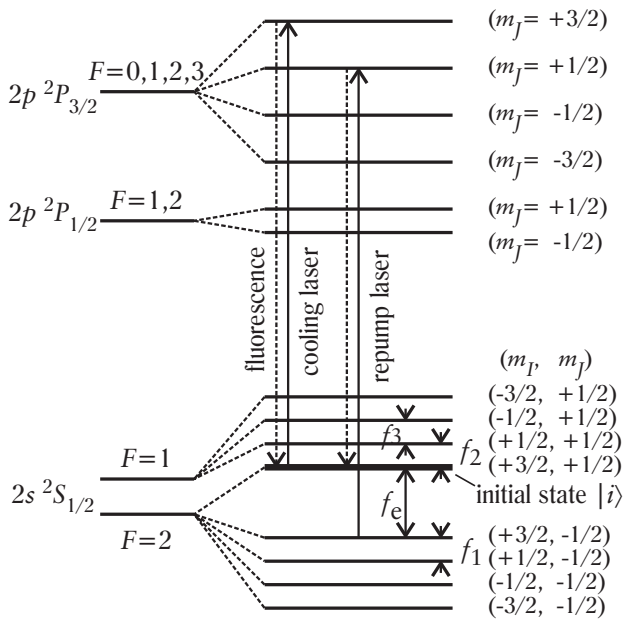}
\caption{Energy level structure of $^9$Be$^+$ at $B \approx$ 4.4609~T. $f_\text{e}\approx 124$ GHz,
$f_1\approx 340$ MHz, $f_2\approx 288$ MHz, and $f_3\approx 287$ MHz. The frequency tuning of the repump laser is shown for fast repumping of the electron spin-flip transition. For the nuclear spin-flip measurements the repump laser was tuned approximately 500~MHz lower than the cooling transition.} \label{fig:EnergyLevelSP}
\end{center}
\end{figure}

  There are three unknown variables ($A$, ${g_I}^\prime/g_J$, and $X$) in
Eq.~(\ref{eq:BreitRabi}), and the measurement of three transition frequencies in the
ground state at a fixed value of $B$ will determine these three variables. 
We experimentally
determined the value of $A$ and ${g_I}^\prime/g_J$ at $B$ $\approx$ 4.4609~T by measuring the three transition frequencies
labeled $f_{\text e}$, $f_1$, and $f_2$ in Fig.~\ref{fig:EnergyLevelSP}.
 While three
frequencies are enough to determine $A$ and ${g_I}^\prime/g_J$, we also measured a fourth frequency $f_3$, to check for consistency.
The typical period required for a complete set of frequency measurements needed to determine $A$, ${g_I}^\prime/g_J$, and $X$ was 30 to 40 minutes. 

We trapped fewer than or approximately 10$^3$~ions in a Penning-Malmberg trap and cooled them
to approximately 1~mK by Doppler laser cooling. The cooling laser also
optically pumped the ions into the ($m_I$, $m_J$)=($\frac{3}{2}$, $\frac{1}{2}$)
state, labeled as the initial state $|i\rangle$ in Fig.~\ref{fig:EnergyLevelSP}. [Here the states are labeled by the ($m_I$, $m_J$) quantum numbers of their largest components.] The
basic experimental procedure for measuring the different transition
frequencies was to (1) turn off the cooling laser, (2) probe the desired transition
with the appropriate rf or microwave radiation, (3) measure the population of
the ions remaining in $|i\rangle$ with the fluorescence induced by the cooling laser,  (4) repump all ions
to $|i\rangle$ with the cooling laser and an additional repumping laser.  We
used the same ions repetitively to measure all transition frequencies.  We first discuss
the basic experimental setup and the 124 GHz
microwave system.  We then discuss in more detail the measurements of the
different transition frequencies and the determination of $A$ and ${g_I}^\prime/g_J$ at high magnetic field.

\subsection{\label{subsec:setup}Experimental setup}

\subsubsection{\label{subsubsec:trap}Penning trap}

\begin{figure}[tbhp]
\begin{center}
\includegraphics[width=8.2cm]{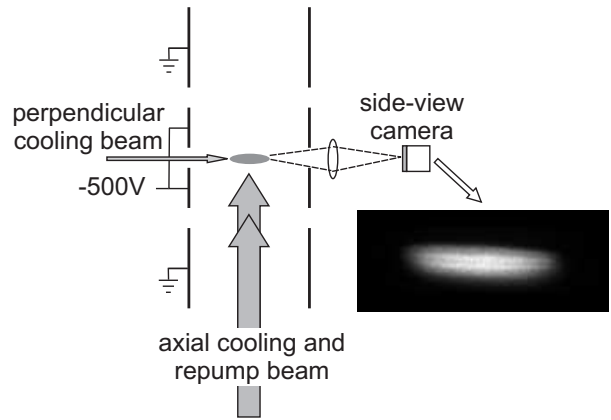}
\caption{Schematic diagram of setup. Figure is not to scale. The
trap diameter is 4~cm. The electrodes used to apply the rotating
wall field are not shown.  The direction of the side-view light
collection and the direction of the perpendicular cooling beam
form a $60^\circ$ angle in a plane perpendicular to the magnetic
field axis.  A side-view image of a plasma with approximately 500 ions is
shown.  The diameter of the fluorescing Be$^+$ ion plasma is
400~$\mu$m.
Heavier-mass impurity ions are located at larger radii than the $^9$Be$^+$ ions \cite{jenm04}.} \label{fig:PenningTrap}
\end{center}
\end{figure}

Figure~\ref{fig:PenningTrap} shows a sketch of the Penning trap
used for the high-$B$ measurements. The trap and the
basic experimental setup have been described
previously~\cite{huap98b,jenm04,biercuk09A}. The 4.4609~T magnetic field of a
superconducting solenoid with a 125 mm room-temperature bore
produces a $^9$Be$^+$ cyclotron frequency of
$\Omega_{c}=2\pi \times 7.602$~MHz. 
The long-term drift of the magnetic field was less than one part in $10^9$ per hour, resulting in an average drift of $f_\text{e}$  of less than 3~kHz per hour. Magnetic field shifts due to changes in the magnetic environment (due, for example, to movement of Dewars or to other activity in neighboring laboratories) were reduced by collecting data during the night.  
We did not stabilize the
pressure in the magnet Dewar or actively cancel external magnetic
field noise, which likely would have improved the long-term magnetic field
stability~\cite{vanr99}. 
The field of the superconducting magnet was found to have fluctuations which were fast compared to 20 Hz, superimposed on the slow magnetic field drift and noise. The frequency spectrum of the fast fluctuations contained a continuous part and narrow peaks  between 30 and 300 Hz \cite{biercuk09A}. The integrated noise of the fast fluctuations produced  $\delta B/B \approx 10^{-9}$ variation in the magnetic field for measurements separated by greater than 0.1~s.  The fast fluctuations contributed to the linewidth and coherence of the electron spin-flip measurement, but had no significant impact on the nuclear spin-flip measurements ($f_1$, $f_2$, and $f_3$).  (The fast fluctuations only produce a phase modulation of a few milliradians on the nuclear spin-flip transitions.)  Recent work, which will be discussed in a separate publication, indicates that the fast fluctuations are fluctuations in the homogeneous field produced by the superconducting magnet, which can be mitigated through vibration isolation of the magnet. 

The Penning trap electrode structure
consists of a stack of four cylindrical electrodes.  The inner
diameters of the cylinders are 4.1 cm and the combined length of the
four cylinders is 12.7 cm. We typically operated the trap with the
central cylindrical electrodes (the ``ring'' electrodes) biased at
$-500$ V and the outer cylindrical electrodes (the ``endcap''
electrodes) grounded, which resulted in $^9$Be$^+$ single-particle
axial and magnetron frequencies of, respectively,
$\omega_{z} = 2\pi \times 565$ kHz and $\omega_{m} = 2\pi \times 21.1$
kHz.

Due to the crossed electric and magnetic fields in a Penning
trap, an ion plasma undergoes a rotation
about the magnetic field axis.  In thermal equilibrium this
rotation is rigid~\cite{dubd99}, and we use $\omega_{r}$ to denote
the plasma rotation frequency.
The rotation frequency $\omega_r$ of the
$^9$Be$^+$ plasma was precisely controlled with a rotating
electric field (a rotating wall)~\cite{huap98b,mitt01}.  A
rotation frequency $\omega_{r}$ of 2$\pi \times$ 30 kHz or less was
used, which produced planar plasmas (oblate spheroids) like that
shown in Fig.~\ref{fig:PenningTrap} with ion densities of
approximately 8 $\times$ 10$^{7}$~cm$^{-3}$. 
The measurements presented here were obtained on small ion plasmas of fewer than 10$^3$ $^9$Be$^+$ ions.
The small axial extent of the
plasmas (typically less than 50 $\mu$m) reduced the effect of
axial gradients in the magnetic field. Axial magnetic field
gradients were shimmed to be less than two parts in 10$^{8}$ per mm,
which resulted in an axial magnetic field inhomogeneity of less
than one part in 10$^{9}$ over a 50 $\mu$m axial extent.  We found
no evidence for any inhomogeneous broadening of the different
resonance curves discussed in Secs.~\ref{subsec:espin} and
\ref{subsec:nucspin}.

\subsubsection{\label{subsubsec:lasers}Laser cooling, state preparation, and detection}

Doppler laser-cooling was carried out on the 313~nm  $^2S_{1/2}\
 (m_I = \frac{3}{2},m_J = \frac{1}{2})$ $\rightarrow$ $^2P_{3/2}\
 (m_I = \frac{3}{2},m_J = \frac{3}{2})$
transition (see Fig.~\ref{fig:EnergyLevelSP}).  The 313~nm light was generated by frequency-doubling the output of a dye
laser at 626~nm.  The axial and perpendicular cooling beams cooled
the motion parallel and perpendicular, respectively, to the
magnetic-field axis \cite{jenm04,jenm05}. The
axial cooling beam had a 1~mm waist diameter, a power of approximately 1 mW, and a
polarization that was either linear or circular ($\sigma^+$).  The
axial cooling beam was aligned with the magnetic field axis to
better than 0.01$^\circ$.  The perpendicular cooling beam was
linearly polarized in a direction perpendicular to $\textbf{B}$,
focused to a waist diameter of approximately 50 $\mu$m, and had a power of approximately 1
$\mu$W. A double-pass acousto-optic modulator was used to rapidly
switch the cooling beams off (in less than 1~$\mu$s) before applying rf or
microwave radiation to drive the desired ground-state transitions.
The cooling beams were switched back on after the rf or microwave radiation was switched off.

The population of the $^2S_{1/2}\ (m_I = \frac{3}{2},m_J = \frac{1}{2})$
state (the initial state $|i\rangle$ in
Fig.~\ref{fig:EnergyLevelSP}) was measured through the
cooling-laser-induced resonance fluorescence. An $f/5$ imaging
system was used to image the $^9$Be$^+$ ion fluorescence onto the
photocathode of a photon-counting imaging tube (quantum efficiency was
approximately 5~\%). The total imaging tube count rate was proportional to
the $|i\rangle$ state population.  The total photon count rate was
recorded for 0.5 s both before and after applying the rf or
microwave radiation.  The ratio of these two count rates (with
small corrections for repumping effects) measured the fraction of
the ions remaining in the $|i\rangle$ state.

The cooling radiation optically pumped more than 94~\% of the ions into the
$^2S_{1/2}\ (\frac{3}{2}, \frac{1}{2})$ state, i.e., the lower level
of the cooling transition \cite{brel88,itaw81}. This was a non-resonant
optically pumping process with a time constant of approximately 5~s for the cooling
laser parameters in this experiment. 
 The repumping time on the electron spin-flip transition ($f_\text{e}$)
was reduced to less than 1~ms by a second
frequency-doubled dye laser (labeled the repump laser in
Fig.~\ref{fig:EnergyLevelSP}). The repump laser was turned on after the second
0.5 s detection period (the detection period after the applied rf or
microwave radiation).  The $f_1$, $f_2$, and $f_3$ transitions involve a change in
the $^9$Be$^+$ nuclear spin orientation.  For example, in $f_2$ the nuclear
spin changes from $m_I = \frac{3}{2}$ to
$m_I = \frac{1}{2}$.  Optical repumping back to $|i\rangle$ occurred
through the $^2S_{1/2}\ (\frac{1}{2}, \frac{1}{2})$  $\rightarrow$
 $^2P_{3/2}\ (\frac{1}{2}, \frac{3}{2})$ transition and the
small admixture of different ($m_I$, $m_J$) states in the $2p\, ^2P_{3/2}$ 
manifold \cite{itaw81}.  We reduced this repumping time
somewhat by tuning the repump laser frequency between 400~MHz and 600~MHz below
that of the cooling transition. This had the added benefit of maintaining a
cold-ion plasma when most of the ions are driven to the
$(m_I = \frac{1}{2}, m_J = \frac{1}{2})$ state.
Presumably this was because the frequency of the repump laser was now below that of the 
$^2S_{1/2}\ (\frac{1}{2}, \frac{1}{2})$ $\rightarrow$
$^2P_{3/2}\ (\frac{1}{2}, \frac{3}{2})$ transition. A similar
improvement in the repumping and plasma stability was also observed for the
$f_1$ and $f_3$ transitions with a ``far-detuned'' laser tuned 400~MHz to 600~MHz
below the cooling transition.  The repump or far-detuned beam was directed
along the magnetic field axis of the trap, as shown in
Fig.~\ref{fig:PenningTrap}. The beam waist diameter was approximately 0.5 mm; the power was a few milliwatts.

\subsubsection{\label{subsubsec:microwave}Microwave apparatus}

\begin{figure}[tbhp]
\begin{center}
\includegraphics[width=8.6cm]{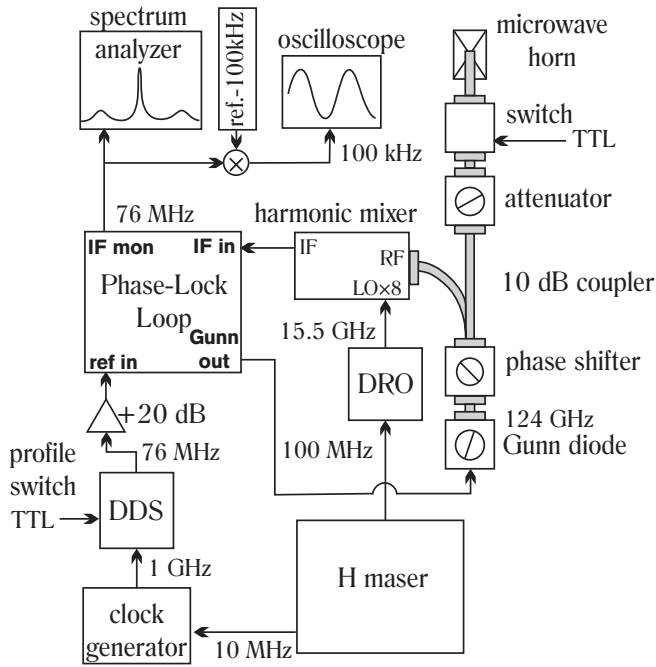}
\caption{Schematic diagram of the 124~GHz microwave source.  A
Gunn diode oscillator generates 30~mW  of microwave power that was transmitted
through WR-8 wave guide (shown in gray) and emitted to free space
via a pyramidal microwave horn. The frequency of the Gunn diode oscillator was 
phase-locked to a signal derived from a hydrogen-maser frequency standard.}
\label{fig:MicroSource}
\end{center}
\end{figure}

A sketch of the 124~GHz microwave system used to measure $f_{\text{e}}$ is
shown in Fig.~\ref{fig:MicroSource}.
Reference \cite{biercuk09A} further discusses the microwave system and its use in quantum information experiments. 
A Gunn diode oscillator generated 30~mW of microwave power, and its
frequency was coarsely set to approximately 124~GHz by a manually tuned microwave cavity.  The
microwave radiation was transmitted through WR-8 wave guides and
launched to free space through a pyramidal rectangular microwave
horn. A small fraction of the microwave power~($-10$~dB) was mixed
with the 8th harmonic of a 15.5~GHz dielectric resonator
oscillator (DRO).  The intermediate frequency (IF) signal from the harmonic mixer was sent
to a phase-locked loop (PLL) controller
and phase-locked to a 76~MHz
reference frequency generated by direct digital synthesis (DDS).
The microwave frequency and phase were controlled by changing the
frequency and phase of the DDS signal. The DDS was controlled by
computer through a parallel interface. All of the frequency
synthesizers used in the
experiment were referenced to the same passive hydrogen maser,
including the DRO and the DDS. The frequency of the passive hydrogen maser was calibrated relative to that of the NIST atomic time scale. 

The microwave radiation emitted from the horn was rapidly switched on and off
with a reflective PIN diode switch.  The ratio of the high and low
power states was $26.5 \pm 0.9$~dB.  Switching between the high- and
low-power states could cause the PLL to lose phase-lock because of the change in the reflected signal. To
avoid losing phase-lock, we added a phase shifter between the Gunn diode
oscillator and the PIN diode switch. (The large fringing field of
the magnet made use of an isolator impractical.) The phase shifter
required careful adjustment to achieve a condition where the Gunn
diode oscillator would not lose lock when the microwave power was
switched.

\begin{figure}[tbhp]
\begin{center}
\includegraphics[width=8cm]{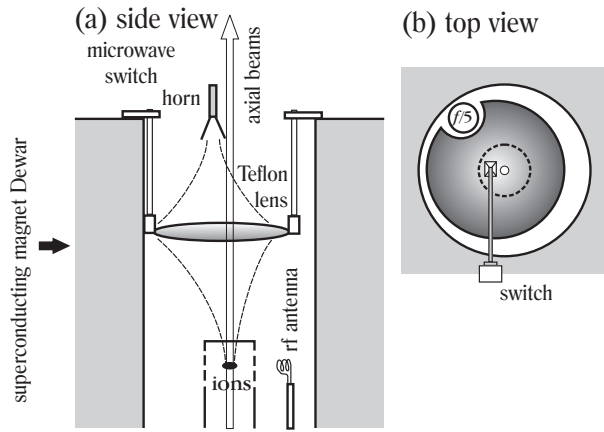}
\caption{Schematic diagram of the quasi-optical coupling of
124~GHz microwave radiation to the ions.  Figures are not to
scale. (a) Side view.  (b) Top view. 
The inner diameter of the super-conducting magnet was
12.5~cm, the distance from the horn to the lens was 29~cm, and from
the lens to the ions is 28~cm. The horn was shifted off-axis in
order to avoid hitting the horn with the axial laser beams.  The
lens was also shifted about half this amount to focus on the ions.
The lens diameter was 10.2~cm. It had a cut on the side
to make a room for the $f/5$ optics used for the side-view camera.
Electron spin-flip $\pi$-pulse periods of 100~$\mu$s were obtained
with this setup. The position of the rf antenna is also shown.}
\label{fig:QuasiOptical}
\end{center}
\end{figure}

We used quasi-optical techniques (with a horn and Teflon\footnote{Teflon is 
a registered trademark of the Dupont Company. Mention of this material should in no way
be construed as indicating that this material is endorsed by NIST or that it is
necessarily the best material for the purpose.} lens) to
couple the microwave radiation to the ions, as schematically shown
in Fig.~\ref{fig:QuasiOptical}.
The pyramidal horn coupled mainly to the Gaussian TEM00 mode with
an initial waist diameter of approximately 0.7 cm.  The waist diameter $w$ is defined by
${P(r)}/{P(0)} = e^{-2(r/w)^2}$, where $P(r)$ is the power per unit area at a
radius $r$.  The waist diameter of the Gaussian beam increases as it
propagates.  The beam was focused to a waist of about 0.7 cm at the ions by the
hyperbolic surfaces of the  lens \cite{golp98}, which was 29~cm below the horn and 28~cm above the ions. The lens had a cut near the side for the $f/5$ side-view
optics and a 4.8~mm diameter hole near center to pass laser
beams~(see Fig.~\ref{fig:QuasiOptical}). The horn was shifted off
axis to avoid the laser beams, and the  lens was shifted
accordingly to center the microwave focus on the ions. The
microwave system was on an $x$-$y$ mechanical stage, and the position of the horn
was adjusted to maximize the coupling of the microwave radiation
with the ions.

Electron spin-flip $\pi$-pulse periods as short as 100~$\mu$s were obtained with this microwave system.  $\pi$-pulse fidelities of better than 99.9~\% were measured with random benchmarking on plasmas consisting of a single plane \cite{biercuk09A}.  The Ramsey free-induction decay (that is, the free-precession period in a Ramsey experiment where the fringe contrast has decayed by 1/$e$) was measured to be $T_2 \approx 2.4$~ms, limited by the fast magnetic field fluctuations.

\subsection{\label{subsec:espin}Electron spin-flip measurement}

\begin{figure}[tbhp]
\begin{center}
\includegraphics[width=8.4cm]{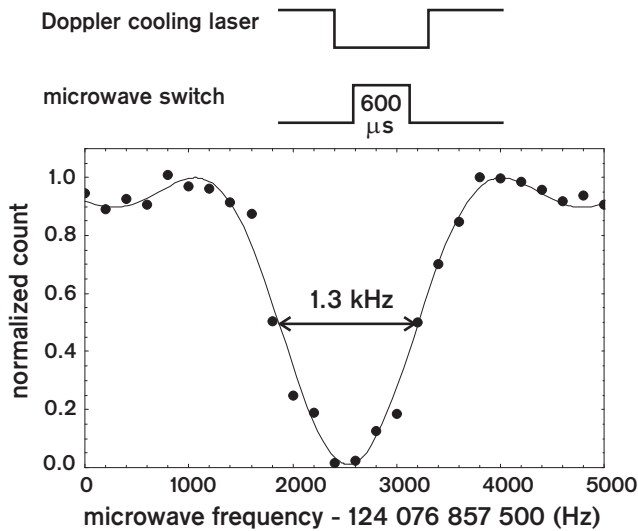}
\caption{Electron spin-flip Rabi resonance. The microwave power is adjusted to produce a $\pi$-pulse on resonance in 600~$\mu$s.
The solid line is a fit to the expected Rabi resonance, as discussed in the text.
All experimental measurements were equally weighted in the fit.} \label{fig:ElectronRabi}
\end{center}
\end{figure}

Figure~\ref{fig:ElectronRabi} shows an electron spin-flip
resonance obtained with a 600~$\mu$s square Rabi pulse.
The data were fitted to the expected
Rabi resonance curve \cite{ramn56},
\begin{equation}
P_i=1-\frac{(2b)^2}{(f_\text{e}-f)^2+(2b)^2}
\sin\left(  \pi t
\sqrt{(f\text{e}-f)^2+(2b)^2}
\right). 
\label{eq:RabiSpectrum}
\end{equation}
Here $P_i$ is the probability of an ion to be in state $|i\rangle$,
$b\equiv\Omega/2\pi$, where $\Omega$ is the Rabi frequency, 
$t=600~\mu$s is the microwave pulse duration, $f$ is the microwave frequency, and $f_\text{e}$ is the electron
spin-flip resonance frequency. From the fit to the data in
Fig.~\ref{fig:ElectronRabi}, we determine a value for the electron
spin-flip frequency $f_\text{e} = 124~076~860~036~\pm 15$~Hz. The
uncertainty obtained from the fit we define to be the internal error, and
for electron spin-flip resonance curves taken under conditions
similar to that shown in Fig.~\ref{fig:ElectronRabi}, the internal
error was typically less than 20~Hz. Because $f_\text{e}$ is roughly
proportional to $B$, 20~Hz corresponds to a
1.6 $\times$ 10$^{-10}$ fractional measurement of $B$, a reduction by about a factor of five, due to averaging, from the shot-to-shot variation in $B$. 

The resonance curve in Fig.~\ref{fig:ElectronRabi} took a few
minutes to obtain.  If we took many resonance curves over a longer
period, the scatter in the fitted values for $f_\text{e}$ was larger
than the internal error of an individual fit, due to slow drift and
fluctuations in $B$. We define the external error to be
the standard deviation of $f_\text{e}$, determined from many separate scans.
The typical external error for data taken over a 30-minute period was greater than
40~Hz.  We found the external error to be smallest between 10~PM and
midnight local time. Since the measurement of $f_\text{e}$ was limited by the
stability of $B$, reducing the internal error by narrowing
the line width with longer Rabi pulses would not have benefited us.

To determine the hyperfine constant $A$, we cycled between measurements of $f_\text{e}$
and measurements of the nuclear spin-flip frequencies discussed in the next
section.  To minimize the effect of drift and slow fluctuations in $B$,
it was important to complete one cycle of measurements as
rapidly as conveniently possible. We found that we could complete a cycle of
measurements more rapidly if we did not change the frequency
of the repump laser to the repumping transition (see
Fig.~\ref{fig:EnergyLevelSP}) for the $f_\text{e}$ measurements. Therefore, we set the
frequency of the repump laser to the far-detuned position (400~MHz to 600~MHz
lower than the cooling transition) for all measurements of $f_\text{e}$, $f_1$, $f_2$,
and $f_3$.

\subsection{\label{subsec:nucspin}Nuclear spin-flip measurements}
The rf radiation used to drive the nuclear spin-flip
transitions was generated by mixing the output of an 80~MHz synthesizer having 
1~mHz resolution with a higher-frequency synthesizer that had lower frequency
resolution. For measurements of $f_2$ and $f_3$ (approximately 290~MHz), the
higher-frequency synthesizer was set to 220~MHz.  For measurements
of $f_1$ (approximately 340~MHz) the higher-frequency synthesizer was set
to 280~MHz. Switching of the rf was done with a switch having
approximately 90~dB isolation.  The rf radiation was coupled to the ions
through a two-turn rf loop antenna, placed near ions, outside
the vacuum envelope (see Fig.~\ref{fig:QuasiOptical}).

\begin{figure}[tbhp]
\begin{center}
\includegraphics[width=8cm]{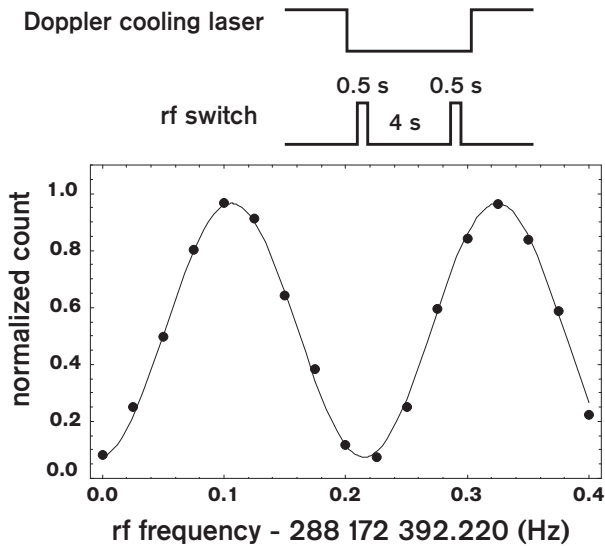}
\caption{$f_2$ resonance obtained with a 4~s Ramsey free-precession
period. The rf power was adjusted to apply a $\pi$/2-pulse in
0.5~s.} \label{fig:F2Ramsey}
\end{center}
\end{figure}

Figure~\ref{fig:F2Ramsey} shows an $f_2$ resonance curve obtained
with the Ramsey method.
The rf power was adjusted to achieve a $\pi/2$ pulse in 0.5~sec.
The two $\pi/2$ pulses were separated by 4~s. After the Ramsey
sequence, both the cooling laser and the far-detuned laser were turned on
simultaneously.  The power of the far-detuned beam was adjusted so
that the dark state repumped with a 1/$e$ time
constant of approximately 2~s. To avoid any significant ac Zeeman shifts due to the
finite isolation (26~dB) of the microwave switch, the microwave
frequency was detuned from resonance with $f_\text{e}$ by 1~MHz during the
$f_2$ measurement. Fitting the data of Fig.~\ref{fig:F2Ramsey} to a
sinusoidal curve gives $f_2$ = 288~172~932.435~3(7)~Hz.  The typical
external error from measurements taken over a 30 to 40 minute measurement cycle
was approximately 5~mHz.  A 5~mHz external error with the 6.5 kHz/mT
 sensitivity of this transition implies a fractional magnetic
field stability of 2 $\times$ 10$^{-10}$ over a typical 30 to 40 minute
period, which is comparable to what was observed on the electron spin-flip
transition.

\begin{figure}[tbhp]
\begin{center}
\includegraphics[width=8.4cm]{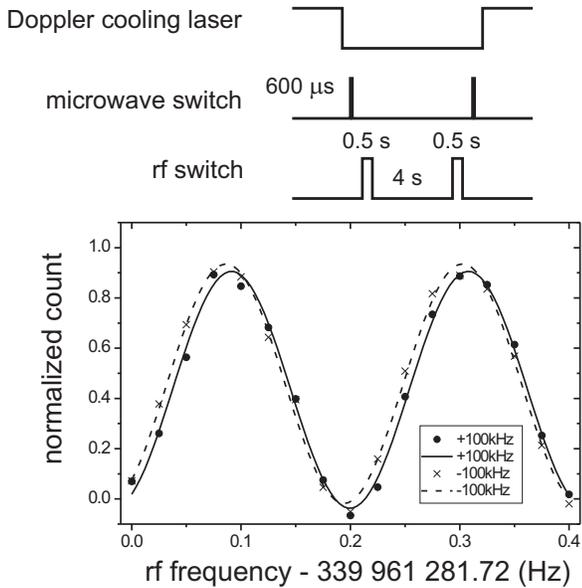}
\caption{An $f_1$ resonance obtained with a 4~s Ramsey free-precession
period. The rf power was adjusted to apply a $\pi$/2-pulse in
0.5~s. The microwave frequency was shifted by $\pm$100~kHz during
Ramsey sequence. } \label{fig:F1Ramsey}
\end{center}
\end{figure}

Figure~\ref{fig:F1Ramsey} shows an $f_1$ resonance curve obtained
with the Ramsey method.
We first transferred the population in $|i\rangle$ to the
($m_I=\frac{3}{2}, m_J=-\frac{1}{2}$) state with a 600 $\mu$s $\pi$-pulse. This
was followed by the Ramsey interrogation, as shown in
Fig.~\ref{fig:F1Ramsey}.  We then used a second 600 $\mu$s
microwave $\pi$-pulse to transfer any ions remaining in the
($m_I=\frac{3}{2}, m_J=-\frac{1}{2}$) state to $|i\rangle$.  The ion population
in $|i\rangle$ was then detected by the laser-induced
fluorescence.

We shifted the microwave frequency by $\pm$100~kHz from resonance
during the Ramsey interrogation of $f_1$. This prevented driving
the $f_\text{e}$ transition with microwave radiation that leaked through
the microwave switch.  The DDS frequency could be switched by as much as 100 kHz, and we could still keep the Gunn diode oscillator phase-locked.
A 100~kHz offset produced a 2~mHz ac Zeeman shift due to the
microwave leakage through the switch. A measurement of $f_1$
consisted of taking two scans with alternate signs of the microwave
detuning. The $f_1$ transition frequency was determined by fitting
the average of the two scans. A fit to the data in
Fig.~\ref{fig:F1Ramsey} provides $f_1$ = 339~961~281.917~0(8)~Hz.  The
external error from measurements taken over a 30 to 40 minute period
was typically 5~mHz, about the same as for the $f_2$ measurements.

\begin{figure}[tbhp]
\begin{center}
\includegraphics[width=8.4cm]{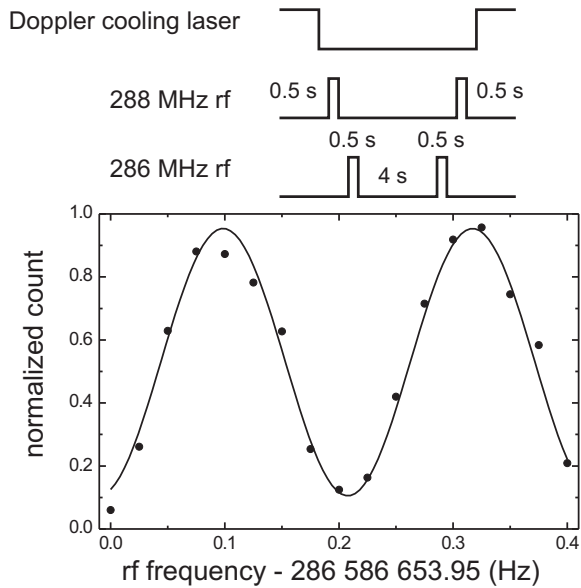}
\caption{An $f_3$ resonance obtained with the Ramsey method.  The power of the rf
was adjusted to apply a $\pi$/2-pulse in 0.5~second.}
\label{fig:F3Ramsey}
\end{center}
\end{figure}

Figure~\ref{fig:F3Ramsey} shows an $f_3$ resonance curve obtained
with a 4~s Ramsey free-precession period.
The pulse sequence was very similar to the one for $f_1$. The
microwave $\pi$-pulse in the $f_1$ measurement was replaced with a
0.5~s rf $\pi$-pulse to transfer the population from $|i\rangle$
to the $(m_I=\frac{1}{2}, m_J=\frac{1}{2})$ state. Then the rf frequency and
amplitude were changed, and we applied a Ramsey sequence with
a 4~s free-precession period. The frequency of the microwave radiation was
detuned from resonance by 1~MHz during the $f_3$ measurement. A fit
to the data in Fig.~\ref{fig:F3Ramsey} gives
$f_3$ = 286~586~654.158~5(14)~Hz.

The length of the Ramsey free-precession periods in the nuclear
spin-flip measurements (4~s) was limited by the heating,
presumably due to collisions with residual gas molecules, that
occurred when the cooling laser was turned off \cite{jenm04}.  For free
precession periods longer than 4~s, the ion fluorescence 
decreased, due to the increase in the Doppler width of the cooling
transition. This added noise and complicated the signal analysis.
Longer free-precession periods (20~s to 100~s) have been used with
sympathetic cooling in previous low-$B$
measurements \cite{bolj89,bolj91}. The nuclear spin-flip
measurements at 4.4609~T were limited by magnetic-field instabilities,
so there was no compelling reason to use longer free-precession periods.

\subsection{\label{subsec:results}Experimental results}

\begin{figure}[tbhp]
\begin{center}
\includegraphics[width=8.6cm]{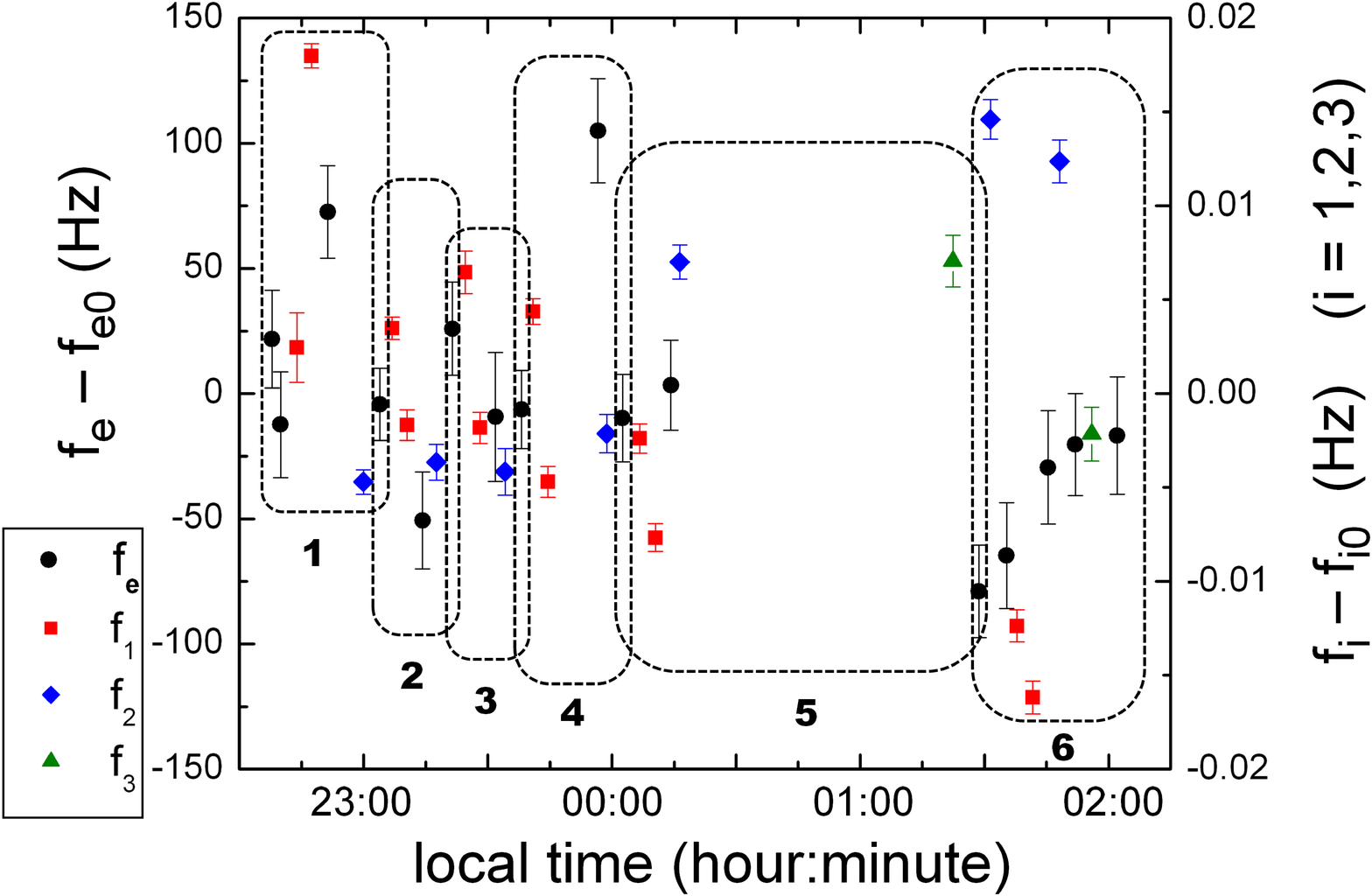}
\caption{
(Color online) Summary of measurements taken during a 4 hour run at night.  The measured electron and nuclear spin-flip frequencies (relative to fixed frequencies $f_{{\text e}0} = 124\,076\,861\,270$ Hz, $f_{10} = 339\,961\,281.922$ Hz, $f_{20} = 288\,172\,932.440$ Hz, and $f_{30} = 286\,586\,654.160$ Hz) are plotted against the time the measurement was made.  The measurements are grouped into cycles, labeled 1--6 on the graph.  Each cycle was used to determine a value of $A$ and ${g_I}^\prime/g_J$.  The error bars are the internal errors obtained from the fits to the resonance curves.  Due to magnetic field drift, $f_{\text e}$ drifted down by about 100 Hz during this run.   This is consistent with the drifts observed with $f_{1}$ and $f_{2}$, where $f_1$ decreases and $f_{2}$ increases with decreasing magnetic field.
}
\label{fig:meascycle}
\end{center}
\end{figure}

To determine the hyperfine constant $A$, resonance
curves such as those shown in
Figs.~\ref{fig:ElectronRabi}--\ref{fig:F1Ramsey} were taken in
succession, as shown in Fig.~\ref{fig:meascycle}, and fits to the resonance curves were used to
determine $f_\text{e}$, $f_1$, and $f_2$. A typical measurement cycle
consisted of an $f_\text{e}$ measurement, followed by an $f_1$ measurement,
followed by another $f_\text{e}$ measurement, followed by an $f_2$
measurement, followed by a final $f_\text{e}$ measurement.  One
measurement cycle took 30 to 40 minutes to complete.  The average of the
three $f_\text{e}$ measurements in one cycle was use to determine $f_\text{e}$. The uncertainty in $f_\text{e}$ was taken to be the external error from the
scatter in the three measurements, which was typically about 40~Hz. The
uncertainties assigned to $f_1$ and $f_2$ were the external errors from the
scatter in the $f_1$ and $f_2$ measurements from consecutive
measurement cycles, which was typically about 5~mHz.

All known systematic errors in the nuclear spin-flip resonance frequency measurements, other than those due to the magnetic field instability, were less than 1~mHz.  The largest systematic error is due to microwave radiation leaking through the microwave switch. As discussed in Sec.~\ref{subsec:nucspin}, this produced a 2~mHz shift in the $f_1$ resonance curve.  However, by taking data with the microwave frequency shifted off resonance by
both $+100$~kHz and $-100$~kHz, this shift could effectively be
canceled. During a 4~s nuclear spin-flip measurement, the ion
temperature increased due to collisions with the room temperature
residual background gas.  Previous studies indicated that the temperature
increase over a 4~s period is limited to a few kelvins \cite{jenm04,jenm05}.
However even a 10~K temperature would produce only an approximately $-0.1$~mHz
time-dilation shift in the measured $^9$Be$^+$ nuclear spin-flip
frequency.  We performed some simple checks for unknown
systematic errors by varying the length of the Rabi pulse in the $f_\text{e}$
measurement and the length of the free-precession period in the
nuclear spin-flip measurements. In addition, we took some $f_1$ and
$f_2$ measurements with an 8~s Rabi pulse. No systematic
dependencies were observed at the level permitted by the magnetic
field stability.

For each measurement cycle, the Breit-Rabi formula
[Eq.~(\ref{eq:BreitRabi})] was used to solve for values of $A$, 
${g_I}^\prime/g_J$, and $X$.  The uncertainties in
these values were determined by using the Breit-Rabi formula to
solve again for $A$, ${g_I}^\prime/g_J$, and $X$, but with $f_\text{e}$, $f_1$, and $f_2$
set to the limits of their uncertainties. We conservatively assigned
the largest uncertainty that could be obtained from the different
combinations of limits. For example, if $\delta f_\text{e}$, $\delta f_1$,
and $\delta f_2$ are the uncertainties in $f_\text{e}$, $f_1$, and $f_2$,
solving the Breit-Rabi formula with the frequency values
$f_\text{e}+\delta f_\text{e}$, $f_1+\delta f_1$, and $f_2+\delta f_2$ gives the
largest uncertainty for $A$, while solving the Breit-Rabi formula
with the frequency values $f_\text{e}+\delta f_\text{e}$, $f_1-\delta f_1$, and
$f_2+\delta f_2$ gives the largest uncertainty for ${g_I}^\prime/g_J$.

\begin{figure}[tbhp]
\begin{center}
\includegraphics[width=8.6cm]{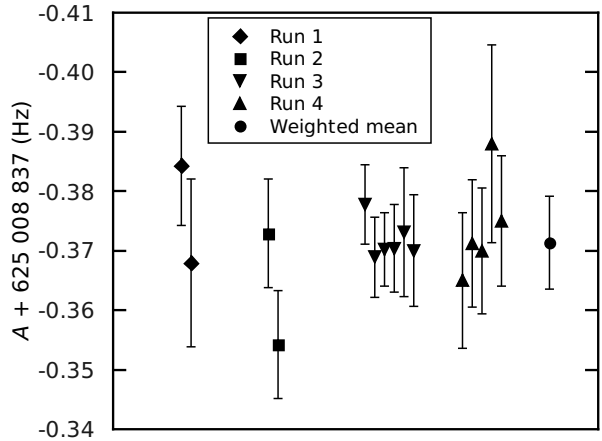}
\caption{Summary of the measurements of the high-field hyperfine
constant $A$.  Each of the four runs is a series of determinations of $A$ made on the same date. Each $A$ value is determined from the Breit-Rabi
formula and one cycle of $f_\text{e}$, $f_1$, and $f_2$ measurements. The
assignment of uncertainties is described in the text.}
\label{fig:Avalues}
\end{center}
\end{figure}

Figure~\ref{fig:Avalues} summarizes the measurements of $A$ at high
magnetic field.  Four different sets of
data were taken on four different dates over a period of two months.
Each set of data consisted of at least two and as many as six
measurement cycles.  The consistency of the data is good.  The
standard deviation from the scatter of the 15 different
measurements of $A$ is 7~mHz, which is slightly less than the average 11~mHz
uncertainty of an individual measurement.  For the determination of $A$
we use a weighted average of the
data shown in Fig.~\ref{fig:Avalues} and conservatively assign an
11~mHz uncertainty, the average uncertainty for a single measurement cycle. 
An 11~mHz uncertainty corresponds to about a $3\times 10^{-10}$ fractional magnetic field instability, about a factor of three below the shot-to-shot fluctuations in the magnetic field. We believe that an assignment of a smaller uncertainty would require a careful study of the statistics of the magnetic field fluctuations. The result is
\begin{equation}
A(4.4609~\text{T}) = -625\,008\,837.371(11)~\text{Hz}.
\label{eq:highB_Avalue}
\end{equation}
Although $A$ was the main focus of this study, a value of the $g$-factor ratio is also obtained from the same analysis:
\begin{equation}
{g_I}^\prime/g_J = 2.134\,779\,852\,7(10) \times 10^{-4}.
\label{eq:highB_gamma}
\end{equation}

In four of the data cycles shown in Fig.~\ref{fig:Avalues}, the fourth frequency $f_3$ was measured in addition to $f_\text{e}$, $f_1$, and $f_2$. 
These measurements were used to place upper limits on corrections to the Breit-Rabi formula at a fixed value of $B$.  The results are listed in Table~\ref{table:f3results}. If the Breit-Rabi formula is assumed to be correct, and $A$ and ${g_I}^\prime/g_J$ are fixed at the values given by the complete set of 15 data cycles [Eqs.~(\ref{eq:highB_Avalue}) and (\ref{eq:highB_gamma})], then any of the three frequencies $f_\text{e}$, $f_1$, or
$f_2$ can be be used to determine $X$ for a given data cycle.  This value of $X$
can then be used to predict $f_3$ for that data cycle.  In practice $f_2$ tended to yield the most consistent values of $X$.
The rms difference between the measured value of $f_3$ and the value predicted from the measurement of $f_2$ was 7~mHz. This is consistent with the noise expected from magnetic field fluctuations. We use these results to place an upper limit of 10~mHz on any shifts of $f_3$ at $B$ = 4.4609~T due to corrections to the Breit-Rabi formula. More specifically, we can set limits on corrections to the Breit-Rabi formula having the form of the quadrupole diamagnetic shift or of the hyperfine-assisted Zeeman shift.
Since all of the measurements were made at nearly the same value of $B$, this test is not sensitive to modifications to the Breit-Rabi formula that amount to a dependence of either $A$ or ${g_I}^\prime/g_J$ on $B$. 

\begin{table*}[tbh]
\caption{Measured values of $f_3$ and values of $f_3$ predicted from the Breit-Rabi formula. All frequencies are in hertz. \label{table:f3results}}
\begin{ruledtabular}
\begin{tabular}{cccc}
$f_2$ (measured) & $f_3$ (predicted from $f_2$) &  $f_3$ (measured) & $f_3$ (predicted) $-$ $f_3$ (measured) \\
\hline
288\,172\,931.651\,0 & 286\,586\,653.404\,9 & 286\,586\,653.400\,3 & 0.0046 \\
288\,172\,931.609\,7 & 286\,586\,653.365\,8 & 286\,586\,653.372\,7 & -0.0069\\
288\,172\,932.447\,0 & 286\,586\,654.156\,8 & 286\,586\,654.167\,1 & -0.0103 \\
288\,172\,932.453\,5 & 286\,586\,654.162\,9 & 286\,586\,654.157\,8 &  0.0051 \\ 
\end{tabular}
\end{ruledtabular}
\end{table*}

The energy shift of an $(m_I, m_J)$ state due to the quadrupole diamagnetic shift, at a fixed value of $B$, has the form \cite{lipson86}
\begin{equation}
E_Q = hf_Q B^2 \frac{[I(I+1)-3m_I^2]}{I(2I-1)}.
\label{eq:quadform}
\end{equation}
If the only correction to the Breit-Rabi formula is given by Eq.~(\ref{eq:quadform}), then agreement of the measured and predicted values of $f_3$ to less than 10~mHz sets a limit $|f_Q|B^2 < 5$~mHz at $B$ = 4.4609~T, or $|f_Q| < 2.5 \times 10^{-4}$~Hz T$^{-2}$.

The hyperfine-assisted Zeeman shift of an $(m_I, m_J)$ state has the form \cite{fortson87}
\begin{equation}
E_{\text{HZ}} = 2h\beta_{\text{HZ}} B [m_I^2 m_J - I(I+1) m_J + m_I/2].
\label{eq:hfszeemanform}
\end{equation}
If the only correction to the Breit-Rabi formula is given by Eq.~(\ref{eq:hfszeemanform}), then agreement of the measured and predicted values of $f_3$ to $\pm 10$~mHz sets a limit 
$|2\beta_{\text{HZ}} B| < 6.7$~mHz at $B$ = 4.4609~T, or $|\beta_{\text{HZ}}| < 7.5 \times 10^{-4}$~Hz T$^{-1}$.

\section{\label{sec:diamagnetic}Determination of $\bm{k}$ from high-field
measurements and previous low-field measurements}

The present experimental results can be combined with previous measurements made by some of the present authors at lower values of $B$ to determine the $B$-dependence of $A$ or ${g_I}^\prime/g_J$. Preliminary values of $A$ and ${g_I}^\prime/g_J$ were given in Ref.~\cite{wineland83} but not the transition frequencies on which they were based. We now supplement Ref.~\cite{wineland83} with the transition frequencies and a final determination of $A$ and ${g_I}^\prime/g_J$.  Two nuclear spin-flip frequencies, labeled 1 and 3 in Table \ref{table:lowfieldfreqs},  were measured near two particular values of $B$ where the first derivatives of the frequencies are zero. Electron spin-flip frequencies, labeled 2 and 4 in Table \ref{table:lowfieldfreqs}, were measured at the same two values of $B$.
The experimental methods have been described in detail \cite{wineland83,bollinger85,bolj89,bolj91}. 
Transition 3 is known better than transition 1 because it was studied for use as a frequency standard \cite{bollinger85,bolj91}.
The value we report for transition 3 in Table~\ref{table:lowfieldfreqs} is slightly different than that reported in \cite{bollinger85} because it includes additional measurements made in 1988 and 1989 as well as an evaluation of the background pressure shift.

If no account is taken of any $B$-dependence of $A$ or ${g_I}^\prime/g_J$, the four frequency measurements, together with the Breit-Rabi formula, yield a system of four equations with four unknowns, $A$, ${g_I}^\prime/g_J$, $X_1$, and $X_2$, where $X_1$ is the value of $X$ for transitions 1 and 2, and $X_2$ is the value of $X$ for transitions 3 and 4.  
The frequencies given in Table \ref{table:lowfieldfreqs} yield $A = -625\,008\,837.053(11)$~Hz and ${g_I}^\prime/g_J = 2.134\,779\,851\,8(23) \times 10^{-4}$.
The precise values of $X_1$ and $X_2$ are not important, since they reflect only the value of $B$ at which the experiment was performed, not any intrinsic property of the $^9$Be$^+$ ion. Comparing these results to Eqs. (\ref{eq:highB_Avalue}) and (\ref{eq:highB_gamma}), we see that there is clear evidence for $B$-dependence of $A$, but that ${g_I}^\prime/g_J$ is independent of $B$ to within experimental error.

\begin{table*}[tb]
\caption{Low-$B$ resonance frequencies used to determine $A$ and ${g_I}^\prime/g_J$.  \label{table:lowfieldfreqs}}
\begin{ruledtabular}
\begin{tabular}{ccccc}
Label & $(m_I,m_J) \leftrightarrow ({m_I}^\prime,{m_J}^\prime)$ & $B$(T) & Frequency (Hz) & Uncertainty (Hz) \\
\hline
1 & $(\frac{1}{2},-\frac{1}{2}) \leftrightarrow (\frac{3}{2}, -\frac{1}{2})$ &  0.677395 & 321\,168\,429.685 & 0.010 \\
2 & $(\frac{3}{2},-\frac{1}{2}) \leftrightarrow (\frac{3}{2}, \frac{1}{2})$ & 0.677395 &  18\,061\,876\,000 & 150\,000 \\
3 & $(-\frac{1}{2},\frac{1}{2}) \leftrightarrow (-\frac{3}{2}, \frac{1}{2})$ &  0.819439 & 303\,016\,377.265\,20 & 0.000\,11 \\
4 & $(-\frac{3}{2},-\frac{1}{2}) \leftrightarrow (-\frac{3}{2}, \frac{1}{2})$ &  0.819439 & 23\,914\,008\,800 & 150\,000 \\ 
\end{tabular}
\end{ruledtabular}
\end{table*}

From theoretical considerations and from experimental results with Rb, a quadratic $B$-dependence of $A$ is expected. If we assume that $A(B) = A_0\times  (1 + kB^2)$, then there are five unknowns to solve for: $A_0$, $k$, ${g_I}^\prime/g_J$, $X_1$, and $X_2$.  
In addition to the four equations for the low magnetic field measurements, derived from the Breit-Rabi formula, a fifth equation is given by the expression for the high-$B$ value of $A$:
\begin{equation}
A_0\times  (1+k{B_3}^2) = -625\,008\,837.371~\text{Hz},
\label{eq:fiftheqn}
\end{equation}
where $B_3 = 4.4609$~T.
Solving the set of five equations gives
\begin{subequations}
\begin{eqnarray}
A_0 & = & -625\,008\,837.044(12)~\text{Hz},\\
k & = & 2.63(18) \times 10^{-11}\;\text{T}^{-2},\\
{g_I}^\prime/g_J & = & 2.134\,779\,852\,0(23) \times 10^{-4}.
\label{eq:finalparameters}
\end{eqnarray}
\end{subequations}
The uncertainties of the parameters are obtained by varying the experimental frequencies through their uncertainties. The value of ${g_I}^\prime/g_J$ given by Eq.~(\ref{eq:finalparameters}) is consistent with, but less precise than, that obtained from the high-$B$ data alone [Eq.~(\ref{eq:highB_gamma})].

\section{\label{sec:theory}Calculation of diamagnetic hyperfine shift coefficient $\bm{k}$}

In nonrelativistic atomic theory, the diamagnetic shift in hyperfine structure arises as a cross term involving both the diamagnetic interaction and the hyperfine interaction in second-order perturbation theory. Let the unperturbed Hamiltonian for an $N$-electron atom with nuclear charge $Ze$ be
\begin{equation}
H_0 = \sum_{i=1}^{N} \frac{\textbf{p}_i^2}{2m} - \sum_{i=1}^{N} \frac{Ze^2}{4\pi\epsilon_0 r_i} 
+ \sum_{i < j}  \frac{e^2}{4\pi\epsilon_0|\textbf{r}_i - \textbf{r}_j|},
\label{eq:unperturbed}
\end{equation}
where $m$ is the electron mass, $-e$ is the electron charge, and $\textbf{r}_i$ and $\textbf{p}_i$ are the position and momentum of the $i$th electron. 
The interaction with an external magnetic field $\textbf{B} = B\hat{\textbf{z}}$ is taken into account by the minimal coupling prescription, i.e., making the replacement $\textbf{p}_i \rightarrow \textbf{p}_i + e\textbf{A}(\textbf{r}_i)$, where $\textbf{A}$ is the vector potential function, $\bm{\nabla} \times \textbf{A} = \textbf{B}$.
Nuclear Zeeman, electron spin Zeeman, and hyperfine interactions are also added as perturbations to $H_0$.
The kinetic energy term for the $i$th electron in Eq.~(\ref{eq:unperturbed}) undergoes the change
\begin{subequations}
\begin{eqnarray}
\frac{\textbf{p}_i^2}{2m} & \rightarrow &\frac{[\textbf{p}_i+e\textbf{A}(\textbf{r}_i)]^2}{2m} \label{eq:minimalcoupling}\\
& = & \frac{\textbf{p}_i^2}{2m} + 
\frac{e[\textbf{p}_i \cdot \textbf{A}(\textbf{r}_i)+ \textbf{A}(\textbf{r}_i) \cdot \textbf{p}_i]}{2m} \nonumber \\
& & + \frac{e^2 \textbf{A}^2(\textbf{r}_i)}{2m}\label{eq:minimalcoupling-b}\\
& = & \frac{\textbf{p}_i^2}{2m} + H_i^\text{p} + H_i^\text{d}.
\label{eq:single-electron-change}
\end{eqnarray}
\end{subequations}
The term containing the first power of $\textbf{A}$ is called the paramagnetic interaction $H_i^\text{p}$, while the one containing $\textbf{A}^2$ is called the diamagnetic interaction $H_i^\text{d}$. The division into paramagnetic and diamagnetic parts is gauge-dependent, but the choice of gauge, 
\begin{equation}
\textbf{A}(\textbf{r})= \frac{1}{2} \textbf{r}\times \textbf{B},
\label{eq:Larmor}
\end{equation}
is particularly convenient. With that choice, the paramagnetic term becomes
\begin{subequations}
\begin{eqnarray}
H_i^\text{p} & = & \frac{e}{2m}[\textbf{p}_i \cdot \textbf{A}(\textbf{r}_i)+ \textbf{A}(\textbf{r}_i) \cdot \textbf{p}_i]\\
 & = & \frac{e}{4m} (\textbf{p}_i \cdot \textbf{r}_i \times \textbf{B} + 
 \textbf{r}_i \times \textbf{B} \cdot \textbf{p}_i)\\
 & = & \frac{e}{2m}\textbf{r}_i \times \textbf{p}_i \cdot \textbf{B}
 = \frac{e}{2m}\bm{\ell}_i \cdot \textbf{B}
 = \frac{e(\ell_i)_z B}{2m},
\label{eq:reduce_paramagnetic}
\end{eqnarray}
\end{subequations}
where $\bm{\ell}_i = \textbf{r}_i \times \textbf{p}_i$ is the orbital angular momentum of the $i$th electron. 
The diamagnetic term becomes
\begin{equation}
H_i^\text{d} = \frac{e^2 \textbf{A}^2(\textbf{r}_i)}{2m}=\frac{e^2}{8m}(\textbf{r}_i \times \textbf{B})^2
= \frac{e^2}{8m}(x_i^2 + y_i^2)B^2.
\label{eq:diamagnetic}
\end{equation}
$H_i^\text{d}$ can be divided into a spherically symmetric (scalar) part and a rank-2 spherical tensor part:
\begin{equation}
H_i^\text{d} = \frac{e^2B^2r_i^2}{12m} + \frac{e^2 B^2 r_i^2}{24m}(1 - 3\cos^2\theta_i)
\equiv H_i^{\text{d}0} + H_i^{\text{d}2}.
\end{equation}

The total paramagnetic interaction is obtained by summing $H_i^\text{p}$ over all electrons:
\begin{equation}
H^\text{p} = \sum_{i=1}^{N} H_i^\text{p} = \frac{eB}{2m}\sum_{i=1}^N (\ell_i)_z
= \frac{eB}{2m}L_z,
\end{equation}
where $L_z$ is the $z$-component of the total electronic orbital angular momentum.
To a very good approximation, the ground electronic state is an $S$-state, that is, an eigenstate of  $\textbf{L}^2$ with $L=0$, so $H^\text{p}$ can be neglected. 

The total scalar part of the diamagnetic interaction is
\begin{equation}
H^{\text{d}0} = \sum_{i=1}^{N} H_i^{\text{d}0} =  \sum_{i=1}^N \frac{e^2B^2r_i^2}{12m}.
\end{equation}
In first-order perturbation theory, this leads to a common shift of all of the hyperfine-Zeeman sublevels of the ground electronic state.  
In second-order perturbation theory, there is a cross term that is first-order in both $H^{\text{d}0}$ and the Fermi contact hyperfine interaction.  This leads to an effective interaction that appears as a shift in the hyperfine $A$ value, proportional to $B^2$ 
\cite{bender64,larson78,economou77,ray79,lipson86A}. Evaluation of the second-order perturbation term yields the constant $k$. 

Unpublished calculations by Lipson, similar to those done for Rb \cite{lipson86,lipson86A}, yielded $k = 2.52 \times 10^{-11}$~T$^{-2}$ for Be$^+$. These calculations used Hartree-Fock-Slater wave functions \cite{herman63}.  The inclusion of all intermediate states, including the continuum, was done by solving an inhomogeneous differential equation for the perturbed wave function \cite{sternheimer51,dalgarno55}.
Similar calculations were done by one of the present authors (W.M.I.), but using a parametric potential for Be$^+$ that reproduces the experimental energy levels \cite{ganas79}.
This yielded $k = 2.68 \times 10^{-11}$~T$^{-2}$. The estimate given in Ref.~\cite{wineland83} of $\Delta A = -0.017$ Hz T$^{-2}$ (equivalent to $k = 2.7 \times 10^{-11}$~T$^{-2}$) was based on these calculations.

Another method of obtaining $k$ to the same order in perturbation theory is to calculate an approximate electronic wave function that is accurate to first order in $H^{\text{d}0}$ and then to calculate the  mean value of the hyperfine interaction with these wave functions. A simple way to do this is to use the MCHF (multiconfiguration Hartree-Fock) or MCDHF (multiconfiguration Dirac-Hartree-Fock) method, where an additional term, $br_i^2$ is added to each single-electron Hamiltonian. Unlike model potential methods, MCHF and MCDHF are {\em ab initio} in the sense that they require no experimental input, such as observed energy levels, only values of fundamental constants. 

The GRASP set of MCDHF programs \cite{grant07,parpia96,jonsson96,jonsson07} were used to calculate correlated wave functions for Be$^+$ with and without the $br_i^2$ term and to calculate hyperfine constants with these wave functions \cite{itano09}. 
The calculation of the unperturbed hyperfine constant $A_0$ for $^9$Be$^+$ was similar to that done by Biero{\' n} et al., \cite{bieron99} but less extensive, leaving out for example nuclear recoil and the Breit interaction.  The result, $A_0 = -624.19$ MHz, is within 0.13~\% of the experimental value. 
The calculation was then modified by including the $br_i^2$ terms in the Hamiltonian. 
In Hartree atomic units ($e = m = \hbar = 1$), $b$ is dimensionless. 
It was varied from $1\times 10^{-5}$ to $2\times 10^{-3}$.  The change in $A$, relative to $A_0$, was found to be proportional to $b$ for $b \leq 1\times 10^{-4}$.
Evaluating the constant of proportionality yields $k = 2.645(2) \times 10^{-11}$~T$^{-2}$, where the uncertainty here reflects only numerical error, not error due to physical approximations, such as neglect of the Breit interaction.  However, given the good agreement of the calculated and experimental values of $A_0$, we estimate the error of the calculated value of $k$ to be no more than 1~\%. All of the calculated values of $k$, including Hartree-Fock-Slater, parametric potential, and MCDHF, are in good agreement with the experimental result within the experimental error of 7~\%.

The use of the diamagnetic  potential in a relativistic calculation requires some justification, since the minimal coupling prescription does not yield a term proportional to $\textbf{A}^2$ in the Dirac Hamiltonian \cite{luber09}. Instead of Eq.~(\ref{eq:minimalcoupling}) we have, for the kinetic energy term in the single-electron Dirac Hamiltonian:
\begin{equation}
c\bm{\alpha} \cdot \textbf{p}_i \rightarrow c\bm{\alpha} \cdot [\textbf{p}_i +e\textbf{A}(\textbf{r}_i)]
= c\bm{\alpha} \cdot \textbf{p}_i + ec\bm{\alpha}\cdot\textbf{A}(\textbf{r}_i),
\label{eq:relmincoupling}
\end{equation}
where 
\begin{equation}
\bm{\alpha} = \left( \begin{array}{cc}
0 & \bm{\sigma} \\
\bm{\sigma} & 0 
\end{array} \right),
\end{equation}
and $\bm{\sigma} = (\sigma^1, \sigma^2, \sigma^3)$ is defined in terms of the Pauli matrices $\sigma^i$.
Since the field-dependent perturbation only contains $B$ to the first power, calculation of $k$ with Eq.~(\ref{eq:relmincoupling}) requires third-order perturbation theory (two orders in the magnetic field interaction and one order in the hyperfine interaction), unlike the nonrelativistic case, which requires only second-order perturbation theory.

Kutzelnigg \cite{kutzelnigg03} showed that a unitary transformation of the Dirac Hamiltonian in the presence of a magnetic field yields terms resembling the nonrelativistic paramagnetic and diamagnetic terms, plus another term which is first-order in $\textbf{A}$ and whose effect goes to zero in the nonrelativistic limit. This justifies the use of second-order perturbation theory to calculate $k$ in the relativistic case.
The relativistic form of the single-electron diamagnetic interaction is \cite{kutzelnigg03,luber09}
\begin{equation}
{H_i^\text{d}}(\text{rel}) = \beta\frac{e^2 \textbf{A}^2(\textbf{r}_i)}{2m}, 
\end{equation}
which differs from the nonrelativistic form [Eq.~(\ref{eq:diamagnetic})] only by the factor of $\beta$,
where $\beta$ is the $4 \times 4$ matrix
\begin{equation}
\beta = \left( \begin{array}{cc}
I & 0 \\
0 & -I 
\end{array} \right),
\end{equation}
where $I$ is a $2 \times 2$ identity matrix. The factor of $\beta$ was found also by Szmytkowski by a different method \cite{szmytkowski02}.
The effect of $\beta$ on a matrix element of ${H_i^\text{d}}(\text{rel})$ is to reverse the sign of the integral involving the product of the small components of the Dirac orbitals.  Thus, the relative error incurred by ignoring $\beta$ should be less than $(Z\alpha)^2$, where $\alpha$ here is the fine-structure constant $e^2/(4\pi\epsilon_0\hbar c)$. This error can be neglected for Be$^+$ ($Z = 4$) but may amount to a few percent for Rb ($Z = 37$).

The total tensor part of the nonrelativistic diamagnetic interaction is
\begin{equation}
H^{\text{d}2} = \sum_{i=1}^{N} H_i^{\text{d}2} =  \sum_{i=1}^N \frac{e^2 B^2 r_i^2}{24m}(1 - 3\cos^2\theta_i) .
\end{equation}
In first-order perturbation theory, this leads to no energy shifts in the ground electronic state.
In second-order perturbation theory, there is a cross term that is first-order in both $H^{\text{d}2}$ and the electric quadrupole hyperfine interaction.  This leads to an effective interaction called the magnetically induced quadrupole hyperfine interaction 
\cite{lipson86,lipson86A}. Comparison of the second-order perturbation expression for the induced quadrupole interaction and the expression for the quadrupole antishielding factor $\gamma_\infty$ defined by Sternheimer \cite{armstrong71} yields
\begin{equation}
f_Q = \frac{e^2 Q \gamma_\infty}{24 mh},
\label{eq:reltogamma_inf}
\end{equation}
where $f_Q$ is the coefficient of the induced quadrupole interaction defined in Eq.~(\ref{eq:quadform}).
This form is useful because values of $\gamma_\infty$ have already been calculated for many atoms and ions.
The value  $\gamma_\infty=0.7088$ for Be$^+$ has been calculated in a Hartree-Fock approximation \cite{langhoff65}. The most recent experimental value of the $^9$Be nuclear quadrupole moment is $5.288 (38) \times 10^{-30}$~m$^2$ \cite{pyykko08}. These values of the constants yield the estimate $f_Q = 6.64 \times 10^{-6}$~ Hz T$^{-2}$, which is much smaller than the experimental upper limit set in Sec.~\ref{subsec:results}. 

The hyperfine-assisted Zeeman shift can be estimated by the same method as that used for the rubidium atom \cite{fortson87}.  In this approximation, the hyperfine matrix elements are given by the Fermi-Segr{\` e} formula \cite{fermi33}, the $ns$-state energies are obtained from a hydrogenic (quantum defect) approximation, and the continuum $s$-state wave functions are obtained from a Coulomb approximation. In this approximation, the coefficient $\beta_{\text{HZ}}$ for $^9$Be$^+$ is equal to $2.61\times 10^{-4}$~Hz T$^{-1}$, which is smaller than the experimental upper limit set in Sec.~\ref{subsec:results}.

Some other $B$-dependent shifts are in principle present, such as a magnetic-field-induced spin-dipole hyperfine term, but based on calculations done for Rb \cite{lipson86A} they are likely to be much smaller than the terms already considered.  Various corrections to $g_J$ and to ${g_I}^\prime$ (e.g., nuclear diamagnetic shielding) are calculable but are beyond the scope of this paper.   For recent calculations of $g_J$ for Be$^+$ and other three-electron atoms, see Refs.~\cite{yan01,yan02,glazov04,moskovkin08}. For a recent calculation of the nuclear diamagnetic shielding of Be$^+$, see Ref.~\cite{pachucki10}.

\begin{acknowledgments}
This article is a contribution of the U. S. government, not subject to U. S. copyright. N. S. was supported by a Department of Defense Multidisciplinary University Research Initiative administered by the Office of Naval Research.
We thank S. J. Lipson for providing an early estimate of $k$ for $^9$Be$^+$.
\end{acknowledgments}


\bibliography{behfsexpt01}

\end{document}